\documentclass[prl,aps,twocolumn,showpacs,superscriptaddress,10pt]{revtex4}

\usepackage{bm}
\usepackage{amsmath,amssymb}
\usepackage{graphicx}
\usepackage{color}
\usepackage{hyperref}
\usepackage[all]{hypcap} 

\newcommand{\mytitle}{Can strong-field ionization prepare attosecond dynamics?}

\newcommand{\rmpdfinfo}{\special{ps:: userdict /pdfmark /cleartomark load put}}

\definecolor{MyDarkGreen}{rgb}{0,0.6,0}
\definecolor{MyDarkBlue}{rgb}{0,0,0.8}
\definecolor{MyDarkRed}{rgb}{0.6,0,0.3}
\hypersetup{breaklinks=true, colorlinks=true,plainpages=true, linktocpage=true, linkcolor=MyDarkBlue, citecolor=MyDarkGreen, urlcolor=MyDarkRed, pdfborder={0 0 0},%
pdfauthor={Stefan Pabst and Hans Jakob W\"orner},%
pdfsubject={Research article},
pdftitle={\mytitle}%
}

\newcommand{\ket}[1]{\left|#1\right>}
\newcommand{\bra}[1]{\left<#1\right|}

\begin{document}

\title{\mytitle} 

\author{Stefan Pabst}
\email[]{stefan.pabst@cfa.harvard.edu}
\affiliation{Center for Free-Electron Laser Science, DESY, Notkestrasse 85, 22607 Hamburg, Germany}
\affiliation{ITAMP, Harvard-Smithsonian Center for Astrophysics, 60 Garden Street, Cambridge, MA 02138, USA}

\author{Hans Jakob W\"orner}
\email[]{woerner@phys.chem.ethz.ch}
\homepage[]{www.atto.ethz.ch}
\affiliation{Laboratorium f\"ur Physikalische Chemie, ETH Z\"urich, Vladimir-Prelog-Weg 2, 8093 Z\"urich, Switzerland}

\date{\today}


\begin{abstract}

Strong-field ionization (SFI) has been shown to prepare wave packets with few-femtosecond periods. Here, we explore whether this technique can be extended to the attosecond time scale. We introduce an intuitive model for predicting the bandwidth of ionic states that can be coherently prepared by SFI. This bandwidth is given by the Fourier-transformed sub-cycle SFI rate and decreases considerably with increasing central wavelength of the ionizing pulse. Many-body calculations based on time-dependent configuration-interaction singles (TDCIS) quantitatively support this result and reveal an additional decrease of the bandwidth as a consequence of channel interactions and non-adiabatic dynamics. Our results further predict that multi-cycle femtosecond pulses can coherently prepare attosecond wave packets with higher selectivity and versatility compared to single-cycle pulses.

\end{abstract} 

\maketitle


The measurement of electronic wave packets has recently attracted widespread interest. Time-domain studies of electrons in atoms, molecules and the condensed phase offer new approaches to understanding electronic structure and electronic correlations (see e.g. \cite{cederbaum99a,mauritsson08a,eckle08a,ott13a,schultze14a,WoLe-PRL-2013,PeHe-PRL-2015}). Electronic wave packets have been measured in the valence-shell of atomic ions using transient absorption \cite{GoKr-Nature-2010,WiGo-Science-2011,ott14a} or sequential double ionization \cite{woerner11a,fleischer11a} and in the valence shell of neutral molecules using high-harmonic spectroscopy (HHS) \cite{kraus13b,baykusheva14a,zhang15a}.

One necessary condition for creating electronic motion is the population of multiple electronic states. Strong-field ionization is well known to fulfill this condition \cite{gibson91a,akagi09a,PaGr-PRA-2012,mcfarland08a,ReHa-PRA-2014}. The second requirement, which has received much less attention, is the creation of coherence between the prepared electronic states. Since ionization is inherently an open-system quantum process with respect to the cation, the coherence between the quantum states of the cation is always imperfect. In other words, SFI leaves the ion in a mixed state which may display no time dependence at all. Hence, a method for predicting the degree of coherence created by SFI is desirable.

Studies based on HHS have so far not considered the role of partial coherence of the transient ionic state \cite{mcfarland08a,smirnova09b,haessler10a,woerner10a,PaSa-JPB-2014}. This situation is largely explained by the fact that, within the current understanding, high-harmonic emission involving multiple electronic states of the cation is insensitive to the coherence between these states as long as ionization and recombination occur to the same electronic state. Hence attosecond electron-hole wave packets discussed e.g. in Refs. \cite{smirnova09b,haessler10a} are obtained under a hypothetical perfect coherence. Sensitivity to electronic coherence only arises in the presence of cross channels that have recently been identified experimentally \cite{kraus13b} and theoretically~\cite{PaSa-PRL-2013,HaBr-PRA-2014}.
The partial coherence of electronic states generated by SFI has however been studied in rare-gas ions, both theoretically \cite{SaDu-PRA-2006,rohringer09a} and experimentally~\cite{GoKr-Nature-2010,WiGo-Science-2011}. These studies showed that the degree of coherence decreases with increasing duration of the ionizing pulse which suggests that the shortest pulses should be used to achieve the highest degree of coherence.

The title question is particularly relevant for applying SFI to initiate charge migration \cite{cederbaum99a,breidbach03a,kraus13c} which is usually discussed in the context of single-photon ionization in the sudden limit. 
A recent study~\cite{PaSa-PRL-2011} of single-photon ionization by attosecond pulses has shown that a necessary condition for the coherent population of cationic states is that the bandwidth of the ionizing radiation exceeds their energetic separation. 
In contrast, the existence of a similar condition for SFI is not obvious. Applying the same rule to SFI would mean that multi-cycle IR pulses can create coherences only between states that are separated by less than the bandwidth of the pulses (typically $<1$~eV), and, therefore, would be unable to prepare attosecond dynamics.
Alternatively the spectral width of the created photoelectron wave packet (2 times the ponderomotive potential) could be the key quantity. This, in turn, would suggest that almost arbitrarily fast dynamics can be prepared. We show that it is in fact a third quantity, although related to the two previous ones, namely the temporal confinement of SFI, that is the key to defining the coherent bandwidth.

In this letter, we introduce the concept of a ''coherence window'' which represents the bandwidth of ionic states that can be coherently prepared by SFI. The coherence window is obtained by Fourier-transforming the sub-cycle time dependence of the strong-field ionization rate. We validate this conceptually intuitive and physically transparent model using the time-dependent configuration-interaction singles (TDCIS) method \cite{GrSa-PRA-2010,Pa-EPJST-2013}, an {\it ab-initio} many-body approach. Our model predicts a pronounced decrease of the coherence window with increasing central wavelength. The TDCIS results reveal a further narrowing of the coherence window caused by channel interactions and non-adiabatic effects. 
Most importantly, all results display energy-domain recurrences of the coherence that enable highly-coherent attosecond wave packets to be selectively prepared by multi-cycle femtosecond pulses. This unexpected property is particularly valuable to achieve selectivity in molecules where SFI would usually prepare highly complex wave packets. 

Our model is motivated by the Fourier principle. The more an event is confined in time, the broader is the associated bandwidth of energies. Applied to SFI, we conjecture that the sub-cycle evolution of the strong-field ionization rate $\Gamma(t):=\Gamma(|{\cal E}(t)|)$ is the key quantity in determining the bandwidth of states that can be coherently prepared, where ${\cal E}(t)$ is the electric-field amplitude. The highly non-linear dependence of $\Gamma$ on ${\cal E}$ results in a wide coherence window that can span several electron volts. We first illustrate this result numerically, then provide an analytical derivation and finally test it against {\it ab initio} TDCIS numerical calculations. 

\begin{figure}[ht!]
  \centering
  \rmpdfinfo
  \includegraphics[clip,width=\linewidth]{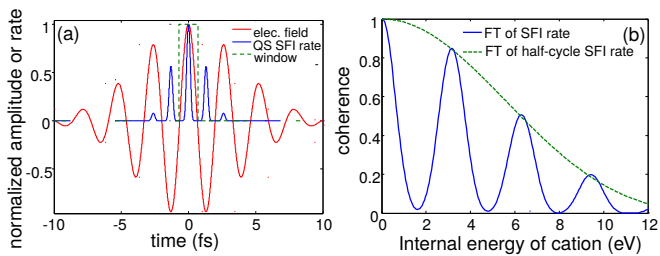}
  \caption{
    (color online) (a) Electric field of a 6.3 fs pulse centered at 800 nm and quasi-static SFI rate for a peak intensity of 10$^{14}$ W/cm$^2$, ionization potential of 12.1 eV and angular momentum quantum numbers $\ell=1$ and $m=0$. (b) Fourier-transform amplitude of the full SFI rate (solid blue) or the SFI rate restricted to the window shown in (a) (green dashed).}
  \label{fig.1}
\end{figure}

Figure \ref{fig.1}a shows a few-cycle pulse and the associated quasi-static SFI rate for a hydrogen-like atom. Figure \ref{fig.1}b shows the Fourier transform of both the complete SFI rate (full blue line) and its restriction to the central half cycle (green dotted line). This latter curve represents the coherence window associated with a single half cycle, whereas the former represents the coherence window associated with the complete pulse. As we show below, the coherence window represents the degree of coherence between the ground state and any excited state of the cation prepared in the SFI process as a function of their energy separation. 

The following conclusions can be drawn from Fig.~\ref{fig.1}, which shows results for 800 nm central wavelength and $10^{14}$ W/cm$^2$ peak intensity. These pulse parameters give rise to a half-width at half maximum (HWHM) of 6.3~eV, corresponding to a temporal period of 0.65~fs. This is the shortest period of a wave packet that can possibly be coherently prepared by SFI under the chosen conditions. As we show below, the width of the coherence window decreases dramatically with central wavelength.
A single-cycle pulse can thus coherently prepare any pair of levels lying within this window (dashed green line in Fig. 1b), whereas a multi-cycle pulse can only coherently prepare levels lying within the narrower peaks of the multi-cycle coherence window (full blue line). These peaks are spaced by 2n$\hbar\omega$ (with $n$ integer) because the SFI rate does not distinguish between positive and negative extrema of the electric field and thus possesses twice the angular frequency $\omega$ of the electric field.

We now provide an analytical derivation of the coherence window. The evolution of the ionic density operator $\hat\rho(t)$ in the interaction picture reads (atomic units are used unless indicated otherwise)
\begin{align}
  \label{eq.1}
  i\partial_t \hat \rho(t)
  &=
  e^{i\hat H_0 t}\, \hat \Gamma(t) \, e^{-i\hat H_0 t}
  ,
  %
\end{align}
where $\hat\Gamma$ describes the ionization rate from the neutral ground state into the singly-ionized states, and $\hat H_0$ describes the field-free propagation of the ionic states. Interaction between the continuum electron and the ion as well as the effect of the strong-field pulse on the ionic states are ignored in Eq.~\eqref{eq.1}. These limitations are removed by turning to TDCIS calculations below. The solution of Eq.~\eqref{eq.1} after the pulse is over ($\hat\Gamma(t')=0, \; \forall t'>t$) is time independent and simply given by 
\begin{align}
  \hat \rho
  \label{eq.2}
  &=
  \sum_{I,J}
    \tilde\Gamma_{IJ}(\Delta E_{IJ})
    \ket{I}\bra{J}
  ,
\end{align}
where $\tilde\Gamma_{IJ}(\nu)=1/2\pi\int\!dt\,\hat\Gamma_{IJ}(t) e^{-i\nu\,t}$ is the Fourier-transformed SFI rate, and $\Delta E_{JI} = E_J - E_I$ is the energy difference between the ionic eigenstates $I$ and $J$. The degree of coherence $C_{IJ}$ between the states $I$ and $J$ of the cation can now be expressed in terms of $\tilde\Gamma_{IJ}$:
\begin{align}
  \label{eq.3}
  C_{IJ}
  &=
  \frac{\big|\rho_{IJ}(t)\big|}{\sqrt{\rho_{II}(t)\,\rho_{JJ}(t)}}
  =
  \frac{\big|\tilde \Gamma_{IJ}(\Delta E_{IJ}) \big|}{\sqrt{\tilde\Gamma_{II}(0)\,\tilde\Gamma_{JJ}(0)}}
  .
\end{align}
If SFI leaves the ion in a pure state, $C_{IJ}=1$, otherwise $C_{IJ} < 1$ with $C_{IJ}=0$ corresponding to a fully incoherent state, which will display no time dependence.

\begin{figure}[ht!]
  \centering
  \rmpdfinfo
  \includegraphics[clip,width=\linewidth]{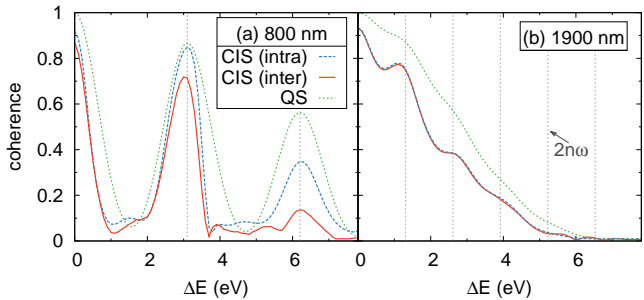}
  \caption{
(color online) Coherence between $5p_{1/2}^{-1}$ and $5p_{3/2}^{-1}$ of Xe$^+$ ($M_J=1/2$) as a function of their separation $\Delta E$ from interchannel TDCIS (red solid), intrachannel TDCIS (blue dashed), and quasi-static (QS) model (green dotted). The ionizing pulse is $T_\text{FWHM}=6.34$~fs long, has a peak intensity of $10^{14}$~W/cm$^2$ and a central wavelength of 800~nm (a) or 1900~nm (b). The vertical dashed lines mark energy splittings corresponding to $\Delta E = 2n\omega$.}
  \label{fig.2}
\end{figure}

We now validate these predictions and additionally analyze the roles played by channel interactions and non-adiabatic effects by turning to TDCIS~\footnote{S. Pabst, L. Greenman, A. Karamatskou, Y.-J. Chen, A. Sytcheva, O. Geffert, R. Santra--\textsc{xcid} program package for multichannel ionization dynamics, DESY, Hamburg, Germany, 2015, Rev. 1349} calculations. As an explicit example, we consider SFI of xenon and study the mutual coherence of the 5p$^{-1}_{1/2}$ and 5p$^{-1}_{3/2}$ states ($M_J=1/2$) of Xe$^+$ as a function of their tunable energy separation $\Delta E$. 
In our calculations the pulse duration as well as the lowest ionization energy are kept constant. Figure \ref{fig.2} shows the coherence $C$ between the two spin-orbit components of the electronic ground state of Xe$^+$ with $M_J=1/2$ following SFI by a 6.3 fs pulse centered at 800~nm or 1900~nm. This pulse duration corresponds to a 2.4-cycle pulse at 800~nm or to a single-cycle pulse at 1900~nm.
Results are shown for quasi-static (QS) tunnelling rates (green dotted) \cite{YuIv-PRA-2001,ToLi-PRA-2005,DeKr-book}, intrachannel TDCIS (blue dashed), and interchannel TDCIS (red solid) calculations. In contrast to the QS model in which the coherence is obtained as the Fourier transform of the SFI rate, the coherence is directly calculated from the results of the TDCIS calculations.
The intrachannel TDCIS model is equivalent to a single-particle picture (for small ground state depletion), in which each ionization channel does not interact with the others. The interchannel TDCIS model additionally includes the interchannel interactions, which can become quite significant in the strong-field regime~\cite{PaSa-PRL-2013}. 

The results obtained at 800~nm (Fig. 2a) all agree in predicting a rapid initial decrease of the coherence with increasing $\Delta E$, followed by two local maxima at $\Delta E = 2n\hbar\omega$. These local maxima are the signature of the multi-cycle nature of the ionizing pulse as discussed above. The predictions from QS, TDCIS (intra) and TDCIS (inter) increasingly deviate toward larger $\Delta E$. 
In contrast, the results obtained for the single-cycle 1900~nm pulse (Fig. 2b) display an almost monotonic decay of the coherence as a function of $\Delta E$. In this case, QS and TDCIS differ much less and the two versions of TDCIS are nearly identical. 

These results are interpreted as follows. First, the intrachannel TDCIS result always lies below the QS result. This shows that the effects neglected in our simple model tend to decrease the coherence window. Second, the interchannel TDCIS result always lies below the intrachannel result, but much more so at 800~nm than at 1900~nm. This shows that interchannel interactions are wavelength dependent and cause more decoherence at shorter wavelengths. Third, the coherence obtained from TDCIS is always less than one. This is attributed to that fact that any hole dependence of the photoelectron (e.g., different energy dependence of the dipole matrix elements) introduces decoherence~\cite{PaSa-PRL-2011,BrPe-book}. In this case, interchannel interactions are again found to cause more decoherence at 800~nm than at 1900~nm.

\begin{figure}[ht!]
  \centering
  \rmpdfinfo
  \includegraphics[clip,width=\linewidth]{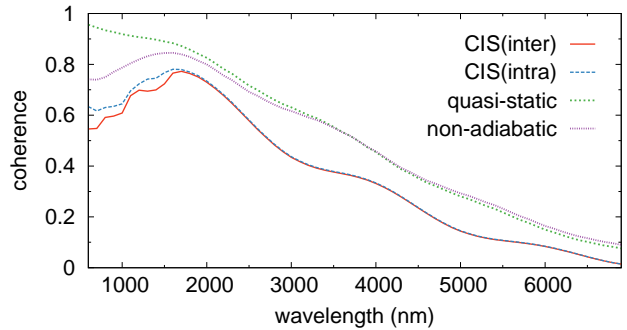}
  \caption{
(color online) Coherence between $5p_{1/2}^{-1}$ and $5p_{3/2}^{-1}$ of Xe$^+$ ($M_J=1/2$ and $\Delta E=1.3$~eV) as a function of the IR wavelength for the interchannel (solid red) and intrachannel (blue dashed) TDCIS, quasi-static (dotted green), and non-adiabatic (dotted purple) models. The pulse is always an one-cycle FWHM pulse with a peak intensity of $10^{14}$~W/cm$^2$.
  }
  \label{fig.3}
\end{figure}

We now study the wavelength scaling of the coherence window and separate the influence of non-adiabatic dynamics from that of the other effects neglected in our simple model. Figure \ref{fig.3} shows the coherence between the two lowest states of Xe$^+$ (using the measured spin-orbit splitting of $\Delta E=1.3$~eV) as a function of the driving wavelength for TDCIS, the quasi-static (QS), and a non-adiabatic (NA) models. The latter consists in Fourier-transforming the non-adiabatic ionization rate derived in Ref.~\cite{YuIv-PRA-2001}. The NA and QS results agree at long wavelengths as expected but show pronounced deviations at shorter wavelengths. The results of the two versions of TDCIS both lie substantially below the QS and NA results for all wavelengths but merge for $\lambda > 2 \mu$m. 

These results show that non-adiabatic effects tend to reduce the coherence generated in SFI. Non-adidabtic effects account for about one half of the coherence reduction from the QS model to the full TDCIS result at 800~nm but much less for $\lambda > 2 \mu$m. The small deviation between the two versions of TDCIS at short wavelengths is attributed to the non-adiabatic interchannel coupling effects which create an entanglement between the photoelectron and the ion that contributes to reducing the coherence within the ion.

\begin{figure}[ht!]
  \centering
  \rmpdfinfo
  \includegraphics[clip,width=.7\linewidth]{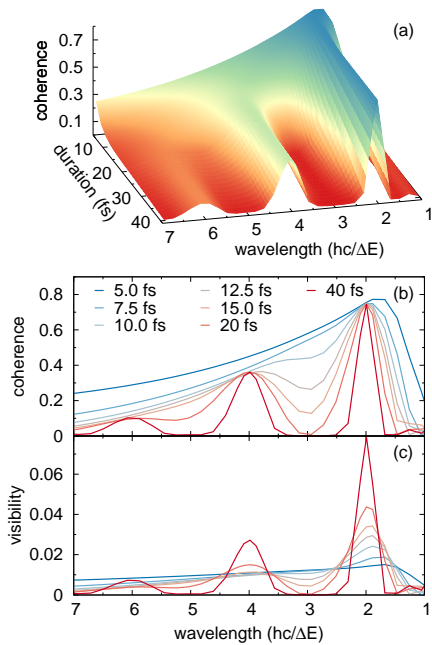}  
  \caption{(color online) 
    (a) Coherence (interchannel TDCIS) between $5p_{1/2}^{-1}$ and $5p_{3/2}^{-1}$ of Xe$^+$ ($M_J=1/2$ and $\Delta E=1.3$~eV) as a function of pulse duration and wavelength in reduced dimensionless units $hc/\Delta E$.
    (b) Cuts for specific pulse durations.
    (c) ``Visibility'' (absolute magnitude of off-diagonal matrix element $|\rho_{IJ}|$) for specific pulse durations.  
    The pulse intensity is fixed at $10^{14}$~W/cm$^2$.
    }
  \label{fig.4}
\end{figure}

The evolution of the coherence as a function of both pulse duration and wavelength in reduced dimensionless units ($\lambda=hc/\Delta E$, for Xe$^+$ i.e. 950~nm) is shown in Fig.~\ref{fig.4}. With increasing pulse duration the peaks in the coherence become narrower, but at the same time the maximal height of each peak remains constant. Together with an increasing ionization probability, this means that the absolute magnitude of the hole oscillations increases with pulse duration (see Fig.~\ref{fig.4}c) and makes the quantum dynamics experimentally more visible. This counter-intuitive result contrasts with the common approaches that try to maximize the hole coherence by using shorter pulses. Our results show that the central wavelength can be a crucial control knob to select pairs of states that are coherently populated and to control the degree of their mutual coherence.

Our results can easily be extended to other atoms, molecules and the condensed matter. By tuning the central frequency of the ionizing pulse coherent hole wave packets between specific ionic states can be created. The common approach of using the shortest possible pulses would in contrast coherently populate all accessible states. Following ionization by a multi-cycle pulse, all accessible states would still be populated but coherence between unwanted pairs of states could be efficiently erased. In addition, the symmetry of the ionic states to be coherently populated is much less important compared to one-photon ionization, where the angular distribution of the photoelectron reveals very strongly the symmetry of the ionic hole (cf. dipole selection rule)~\cite{PaSa-PRL-2011}.

Our discussion of electronic hole coherence applies equally to vibronic (electronic and vibrational) states. In reality this means, however, for molecules and solids, the dephasing due to electron-phonon coupling sets an upper bound to the pulse duration, since all the vibronic states for a specific electronic character need to fit within one coherence peak.

Finally, we use this new knowledge to propose two-dimensional attosecond quantum-beat spectroscopy which works as follows. A few- to multi-cycle IR pulse is split into two replicas that are delayed with respect to each other. The total double-ionization yield is monitored as a function of delay and is, then, Fourier transformed. If the system under study possesses excited states lying within the coherence window, the first pulse will prepare a highly-coherent wave packet that will modulate the probability of the second ionization step \cite{woerner11a,fleischer11a}. Such measurements are repeated for a range of central wavelengths of the pulses. 
The advantages of this technique compared to current approaches are the following: (i) subfemtosecond resolution is achieved with multi-cycle, wavelength-tunable pulses that are readily available from commercial laser systems, (ii) there are no ''dark'' states in SFI as opposed to methods relying on one-photon transitions such as fluorescence or absorption, and (iii) the sub-cycle evolution of molecular SFI can be projected to the energy domain where it is much more accessible. This last property may offer new approaches to studying SFI rates on the sub-cycle time scale.

In conclusion, we have introduced an intuitive approach to predicting the degree of coherence between multiple states of a cation prepared by SFI. We showed that non-adiabatic effects and channel interactions generally tend to decrease the degree of coherence predicted by this simple model.
We have further shown how coherent hole wave packets can be selectively created with multi-cycle strong-field pulses that may be much longer than the period of the hole dynamics itself. This approach does not require single- or few-cycle IR pulses that are challenging to create. The hole coherence maximizes when the relation $\Delta E = 2n\hbar\omega$ is fulfilled, which reflects the situation where each half-cycle of the pulse creates a hole that is in phase with the one created in earlier half cycles. 
The described approach also offers a way to selectively create coherent wave packets involving specific hole states that could not be generated with short ``delta-like'' strong-field or attosecond XUV pulses. A generalization to multi-color pulses opens the path to selectively create even more complex hole wave packets involving several ionic states.

\begin{acknowledgments}
S.P. is funded by the Alexander von Humboldt Foundation and by the NSF through a grant to ITAMP.
S.P. thanks the Helmholtz association for financial support through the Helmholtz Ph.D. prize. 
H. J. W. gratefully acknowledges funding from the European Research Council through an ERC starting grant (contract No. 307270-ATTOSCOPE).
\end{acknowledgments}

\bibliographystyle{apsrev4-1}
\bibliography{amo,books,attobib}


\end{document}